\documentclass[prd,12pt,superscriptaddress,nofootinbib]{revtex4-1}
\usepackage{amssymb}
\usepackage{graphicx,setspace}
\usepackage{amsmath}
\usepackage{color}
\usepackage{mathrsfs}
\usepackage{hyperref}

\begin{document}
\title{Some aspects of dispersive horizons: lessons from surface waves}
\author{J. Chaline}
\affiliation{Universit\'{e} de Nice-Sophia Antipolis, Laboratoire J.-A. Dieudonn\'{e}, UMR CNRS-UNS 6621, Parc Valrose, 06108 Nice Cedex 02, France, European Union.}
\author{G. Jannes}
\affiliation{Universit\'{e} de Nice-Sophia Antipolis, Laboratoire J.-A. Dieudonn\'{e}, UMR CNRS-UNS 6621, Parc Valrose, 06108 Nice Cedex 02, France, European Union.}
\affiliation{Low Temperature Laboratory, Aalto University School of Science, PO Box 15100, 00076 Aalto, Finland, European Union.}
\author{P. Ma\"{i}ssa}
\affiliation{Universit\'{e} de Nice-Sophia Antipolis, Laboratoire J.-A. Dieudonn\'{e}, UMR CNRS-UNS 6621, Parc Valrose, 06108 Nice Cedex 02, France, European Union.}
\author{G. Rousseaux}
\affiliation{Universit\'{e} de Nice-Sophia Antipolis, Laboratoire J.-A. Dieudonn\'{e}, UMR CNRS-UNS 6621, Parc Valrose, 06108 Nice Cedex 02, France, European Union.}

\begin{abstract}
{\bf Hydrodynamic surface waves propagating on a moving background flow experience an effective curved space-time. We discuss experiments with gravity waves and capillary-gravity waves in which we study hydrodynamic black/white-hole horizons and the possibility of penetrating across them. Such possibility of penetration is due to the interaction with an additional ``blue'' horizon, which results from the inclusion of surface tension in the low-frequency gravity-wave theory. This interaction leads to a dispersive cusp beyond which both horizons completely disappear. We speculate the appearance of high-frequency ``superluminal'' corrections to be a universal characteristic of analogue gravity systems, and discuss their relevance for the trans-Planckian problem. We also discuss the role of Airy interference in hybridising the incoming waves with the flowing background (the effective spacetime) and blurring the position of the black/white-hole horizon.
}

\end{abstract}

\maketitle

\section{Introduction}
Several physical systems reproduce certain properties of astrophysical objects like black holes: they exhibit an effective curved space-time when a wave propagates in a ``moving'' medium~\cite{BLV, SUbook}. Examples can be found in acoustics, dielectrics, optical fibres, micro-wave guides, Bose-Einstein condensates, superfluids, ion traps...
Starting with the seminal works of White~\cite{White}, Anderson \& Spiegel~\cite{AS}, Moncrief~\cite{Moncrief} and Unruh~\cite{Unruh}, there has been a growing interest for such analogue models of gravity in order to simulate and understand the physics of wave propagation on a curved space-time. The case of interface and surface waves was initiated by Sch\"utzhold \& Unruh, who derived the equation of propagation of long gravity waves moving on a background flow in terms of a general relativistic metric~\cite{SU}. A black hole, or its time-inverse: a white hole, can indeed be mimicked by the interaction between interfacial waves and a liquid current~\cite{SU, NJP08, PRL09, NJP10, Silke}. Just like the expected behaviour of light near the event horizon of a black hole, gravity waves cannot escape a hydrodynamic black hole featuring a trapping line caused by the velocity gradient of a sufficiently strong background flow. In hydrodynamic experiments, it is usually more convenient to simulate a white hole. Then, gravity waves cannot enter a hydrodynamic white hole featuring a blocking line. 

In a previous series of experiments, such an artificial white hole was created and observed in a laboratory, using water waves  in the presence of a counter-flow inside a wave-tank with varying bottom profile~\cite{NJP08}. Here, we provide a status report on a new series of ongoing experiments, in which we study the behaviour of surface waves of various frequencies in the gravity and gravity-capillary regime in the presence of such white-hole horizons. We encounter a good qualitative agreement with the theory developed in~\cite{PRL09, NJP10}, and interpret the interesting quantitative differences. We are in particular interested in the following key points of the theory of gravity-capillary surface waves. When including surface tension, a second and third horizon appear on top of the white-hole horizon for gravity waves: a ``blue'' horizon associated to mode-converted blue-shifted waves, and a ``negative'' horizon. This negative horizon is associated to the appearance of negative-energy waves~\cite{NJP08} (recently observed in a similar setup~\cite{Silke}), an essential feature of Hawking radiation (the quantum glow of super-massive objects). The blue horizon can interact and even merge with the white horizon providing two scenarios in order to escape an artificial black hole. One scenario, consisting of a double bounce with mode conversion, was 
experimentally discovered nearly three decades ago by Badulin et al.\ 
\cite{Badulin}, and interpreted in terms of the black/white-hole analogy in 
\cite{NJP10}. The other scenario is novel and consists of a direct dispersive 
penetration across the white-hole horizon. We have observed strong indications 
to validate this second scenario, and hope to confirm these quantitatively in 
the near future. This would establish that one can enter into the (normally forbidden) white-hole region by sending high-frequency capillary waves. These are not blocked at the primary white-hole horizon, contrarily to the low-frequency gravity waves, nor even at higher counter-flow velocities.\footnote{In the presence of an ever-increasing counter-flow velocity, the capillary waves will continuously blueshift and ultimately vanish through viscous damping. We will come back to this point in section~\ref{S:transplanckian}.} Indeed, the horizon completely disappears above a certain frequency due to the merging and consequent disappearance of the blue and white horizons.

These scenarios are not limited to hydrodynamic models, but are generic consequences of the dispersive properties beyond the relativistic regime. A double-bouncing scenario is possible in any system where ``subluminal'' dispersion (group velocity $c_g$ decreasing with the wave number $k$) at intermediate wavenumbers $k$ gives place to ``superluminal'' dispersion ($c_g$ increasing with $k$) at higher $k$. This latter condition (superluminal high-$k$ dispersion, irrespective of the behaviour at intermediate $k$) alone is sufficient for direct dispersive penetration. 

\section{Preliminaries}
Water waves in the presence of a uniform current are described by the dispersion relation~\cite{FS, Badulin, Dingemans, Huang}
\begin{equation}\label{disp-rel-general}
(\omega -{\bf U.k})^2 = \left( gk+\frac{\gamma}{\rho} k^3 \right) \tanh(kh)
\end{equation}
where $\omega /2\pi$ is the frequency of the wave in the rest frame and $k$ the wavenumber; $g$ denotes the gravitational acceleration at the water surface, $\rho$ the fluid density, $\gamma$ the surface tension, $U<0$ the constant velocity of the background flow and $h$ the water depth. For water, $\rho = 1000 \mathrm{kg.m^{-3}}$ and $\gamma = 0.073 \mathrm{N.m^{-1}}$. The flow induces a Doppler shift of the pulsation $\omega$.

The relativistic regime with its Schwarzschild-like metric corresponds to the shallow-water limit ($kh\ll 1$) of gravity-wave propagation~\cite{SU}: $(\omega -{\bf U.k})^2 \simeq ghk^2$, with the relativistic ``invariant'' speed $c=\sqrt{gh}$. An analogue black/white-hole horizon will then appear when $|U|=\sqrt{gh}$. But the concept of horizon can easily be generalized to any location where the group velocity $c_g\equiv d\omega/dk$ vanishes. The shallow-water limit $kh\ll 1$ can be realized for example in the circular hydraulic jump, which spontaneously forms a hydrodynamic white hole~\cite{Jannes:2010sa}. In wave-channel experiments, both the shallow-water and the deep-water limits can be probed, depending on the setup. Interaction with a counter-current will lead to a blueshifting of the incident waves: $k$ increases. But variations of the counter-current flow rate are typically achieved by a variation of $h$ (e.g. through the immersion of a bump). These two effects compete and various $kh$ regimes can in principle be realised. At sufficiently high wavenumbers $k$, small-scale dispersive corrections appear due to capillarity, which will turn out to be crucial to cross a horizon. Since this is our main object of study, we will from now on focus on the deep-water limit ($kh\gg 1$), and start again from long-wavelength gravity waves, proceeding step by step towards higher $k$.

When a gravity wave meets a counter-current, the incident wavelength diminishes and the wave height increases~\cite{FS, Badulin, Dingemans, Peregrine, Smith, PS, Basovich, Chawla1, Chawla2, Suastika1, Suastika2, Igor, Baschek}. According to the ray theory, the wave amplitude would diverge when blocking occurs. However, such caustic for the energy is avoided by a regularization process. Due to the velocity gradient, the incoming waves are somewhat diffracted before being stopped at the blocking point, where $c_g$ changes sign and the waves are reflected. Blue-shifted modes are therefore created through a process of mode conversion at the blocking point. Since the incident and blue-shifted waves have the same wavenumber at the blocking point, they interfere and a spatial resonance appears. The diffraction implies that the figure of interference is not a simple standing wave. In the deep-water limit valid for our experiments, an Airy interferences pattern appears~\cite{PRL09, Smith, PS, Basovich}. The blocking point itself is a saddle-node or tangent bifurcation~\cite{PRL09}: it marks the point where the two real solutions disappear. 

\begin{figure}[!htbp]
\begin{center}
\includegraphics[width=0.65\textwidth]{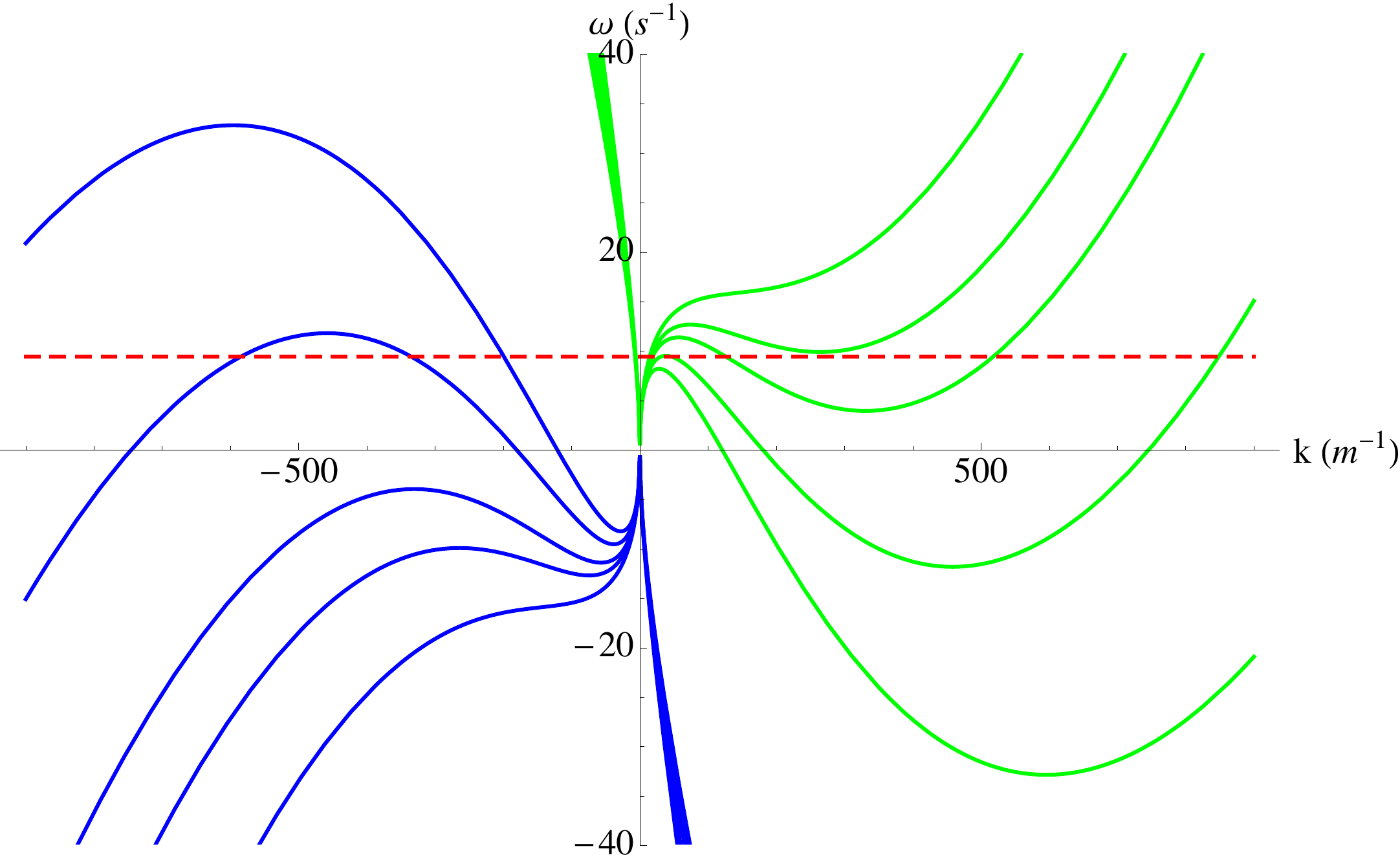}
\caption{Dispersion relation~\eqref{cubic-dispersion} for varying counter-flow $U$ (the curves rotate clockwise with increasing $|U|$), plotted in the form $\omega=Uk\pm \sqrt{\left(gk+\frac{\gamma}{\rho} k^3 \right)\tanh(kh)}$. The group velocity $c_g=\frac{d\omega}{dk}$ corresponds to the slope of the green/blue curves (green: positive co-moving frequency $\omega'=\omega -{\bf U.k}$, blue: negative $\omega'$), and horizons are characterized by local minima/maxima. The double-bouncing observed by Badulin~\cite{Badulin} corresponds to the dashed red line (see also~\cite{NJP10}).}
\label{Fig:disp-rel}
\end{center}
\end{figure}

When including surface tension~\cite{NJP10}, the deep-water ($kh\gg 1$) dispersion relation becomes
\begin{equation}\label{cubic-dispersion}
(\omega -{\bf Uk})^2 \simeq gk+\frac{\gamma}{\rho} k^3,
\end{equation}
or $(\omega -{\bf Uk})^2 \simeq gk \left(1+l_c^2k^2\right)$, where $l_c \equiv \sqrt{\frac{\gamma }{\rho g}}$ is the capillary length ($l_c\simeq 2.7$mm for water). 
The gravity waves are still blocked, with the blocking velocity for pure gravity waves ($\gamma=0$) given by $|U_g|=gT/8\pi$, with $T=2\pi/\omega$ the period. Moreover, the blue-shifted waves are also stopped at a new blocking point on their backward drift (group velocity $c_g<0$, along with the background flow $U$, although the phase velocity $c_{\phi}\equiv \omega/k>0$). The 
capillary asymptotic limit for the blocking velocity of these blue-shifted waves is $U^*_{T\to \infty}=U_\gamma=-\sqrt{2}\left(\frac{\gamma g}{\rho}\right)^{1/4}$~\cite{NJP10}. At this second blocking point, the blue-shifted wave merges with a new capillary solution, which appears (again through mode conversion, see Fig.~\ref{Fig:disp-rel}) at this secondary turning point. The capillary waves propagate in the same direction as the original incident gravity waves. These newly created capillary waves are not blocked by the primary saddle-node line but go through the gravity horizon~\cite{NJP10}. Badulin et al.\ observed experimentally that gravity waves can undergo such double bouncing behaviour followed by conversion to capillary waves which propagate into the forbidden region, and finally vanish by viscous damping~\cite{Badulin}. Fig.~\ref{Fig:disp-rel} illustrates the process.
In the context of the black hole analogy, when time-reversing these observations, one concludes that incident long-wavelength gravity waves cannot escape from a trapping region (black hole) unless they are converted into the capillary range. 

\begin{figure}[!htbp]
\begin{center}
\includegraphics[width=16cm]{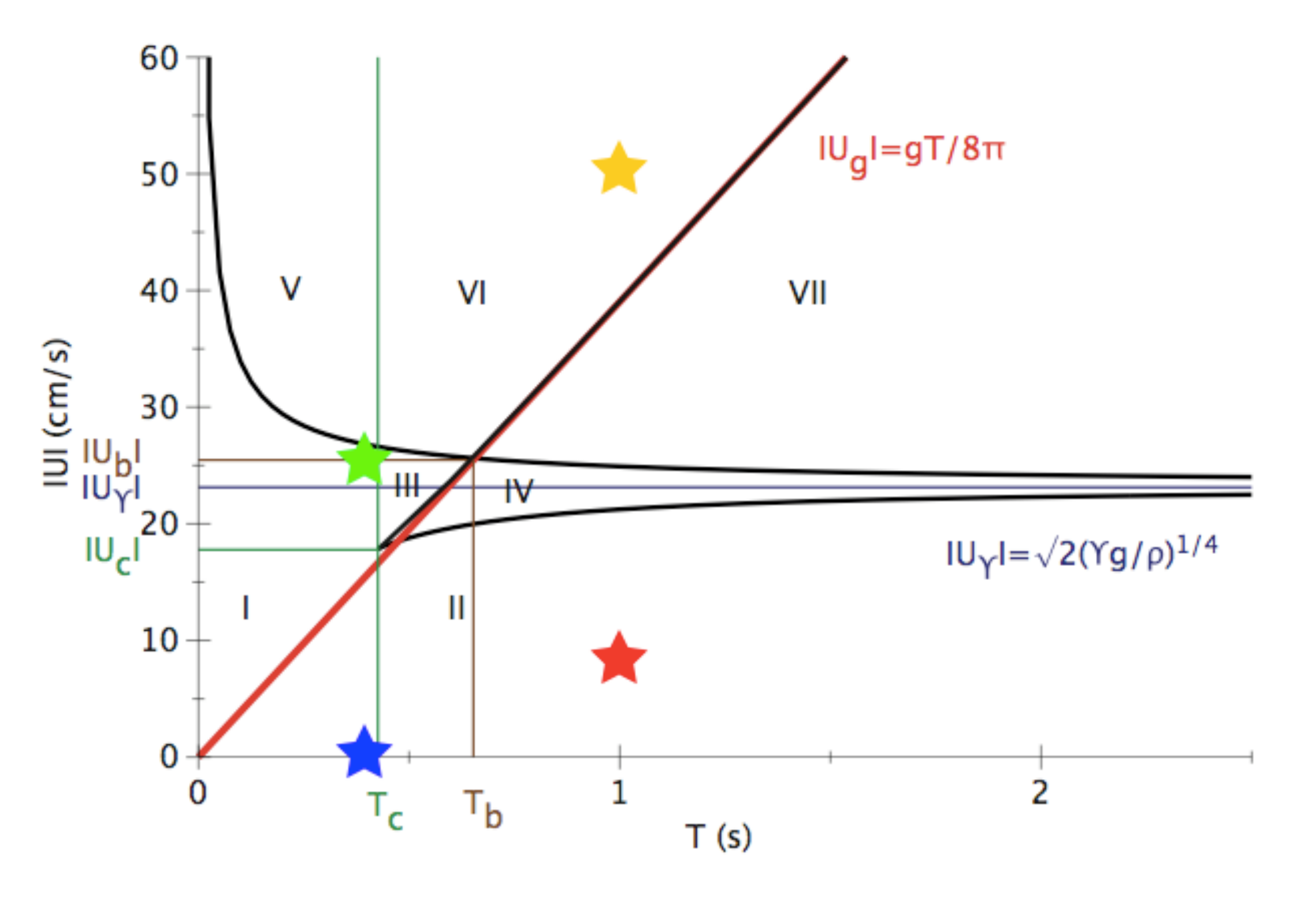}
\caption{Phase space: background counter-flow velocities $U$ versus wave period $T$. The blocking curves $U^*$ are marked in thick black lines. The white-hole horizon (corresponding to the gravity-wave blocking curve $U_g$---thick red line---when $\gamma=0$) intersects the blue horizon or blocking curve for the blueshifted waves (the lowest thick black line, which asymptotes to $U_\gamma$ from below) and creates a cusp $(T_c, U_c)$ below which both horizons disappear. Note that there also exists a blocking line for the negative-frequency waves, which intersects $U_g$ at $(T_b,U_b)$ and also asymptotes to $U_\gamma$ (from above). See also~\cite{NJP10}. The four coloured stars correspond to the experimental results reported in Section~\ref{S:results}.}
\label{Fig:phase}
\end{center}
\end{figure}

The second escape route consists of creating incident waves from the start in the capillary range. 
%
The horizon completely disappears below some critical period $T_c$, determined by the cusp formed through the interaction of the white and blue horizons. Waves with $T<T_c$ can then propagate straight ahead, avoiding any horizons, and enter the ``forbidden'' region (or escape from the trapping region). The cusp can clearly be identified graphically from the ($U^*$ vs $T$) phase diagram for deep-water waves ($kh\gg 1$), see Fig.~\ref{Fig:phase}, where $U^*$ represents any critical or blocking velocity. 

The theoretical phase diagram Fig.~\ref{Fig:phase} was derived in~\cite{NJP10}. We briefly recall the main steps in its derivation. We start from the cubic dispersion relation~\eqref{cubic-dispersion}
for water waves: $(\omega -Uk)^2 \simeq gk+\frac{\gamma}{\rho} k^3 $, where we have taken $k>0$. A double root $k_2$ of this cubic equation, characteristic of a saddle-node point (and hence a horizon or turning point), is such that $(k-k_1)(k-k_2)^2=0$, where $k_1$ is the remaining simple root. 

After some straightforward but tedious algebra, this constraint leads to a quintic equation for the critical velocity $U^*$, corresponding to all possible resonances (for $k>0$):
\begin{equation}\label{quintic}
4\rho ^2 g \omega [ U^5+\frac{1}{4}\frac{g}{\omega}U^4+\frac{\gamma \omega ^2 }{\rho g}U^3-\frac{15}{2}\frac{\gamma \omega }{\rho }U^2-6\frac{g\gamma  }{\rho }U\\
-(\frac{\gamma g^2 }{\rho \omega}+\frac{27}{4}\frac{\gamma ^2 \omega ^3 }{\rho ^2 g}) ]=0
\end{equation}
A dual quintic is obtained for $k<0$ by reversing the velocity $U\rightarrow -U$.

Using this quintic, one can numerically compute and plot the velocity for saddle-node bifurcations or blocking points as a function of the period of the incident waves, as in Fig.~\ref{Fig:phase}. Here we are mainly concerned with the gravity-wave blocking speed $U_g$ and the blocking speed for blue-shifted waves (of which $U_\gamma$ is the asymptotic limit as $T\to \infty$), but we note that there also exists a blocking line for the negative-frequency waves~\cite{NJP10}.

The well-known result $U_g=-g/4\omega$ in the gravity-wave limit immediately follows from eq.~\eqref{quintic} when setting $\gamma=0$ (or $|U|\to \infty$). The asymptotic capillary limit $U_\gamma=-\sqrt{2}\left(\frac{\gamma g}{\rho}\right)^{1/4}$ is likewise obtained for $T=\frac{2\pi}{\omega} \to \infty$. From eq.~\eqref{quintic}, several analytic approximations can also be obtained. For example, to find an approximate expression for the blue horizon (the blocking of blue-shifted waves), we keep the terms up to first order in $\omega$
and introduce the perturbative development
\begin{equation}
U^*\simeq U_\gamma \left(1+\frac{\epsilon}{U_\gamma}\right)
\end{equation}
in order to write:
\begin{equation}
4{\rho\omega}U_\gamma ^5 \left(1+\frac{5\epsilon}{U_\gamma }\right) +g{\rho}U_\gamma ^4 \left(1+\frac{4\epsilon}{U_\gamma }\right)-24g{\omega\gamma}U_\gamma \left(1+\frac{\epsilon}{U_\gamma}\right)-4g^2\gamma \simeq 0.
\end{equation}
Solving for $\epsilon$, using $\omega \to 0$, we obtain 
\begin{equation}
\epsilon \simeq \omega \left(\frac{6\gamma}{\rho U_\gamma ^2}-\frac{U_\gamma ^2}{g}\right)
\end{equation}
or
\begin{equation}
U^*\simeq U_\gamma  + \frac{2\pi}{T}\sqrt{\frac{\gamma}{g\rho}} = - \sqrt{2}\left(\frac{g\gamma}{\rho}\right)^{1/4}+ 2\pi\frac{l_c}{T} .
\end{equation}
Note that an identical computation for the negative quintic leads to a simple change of sign in the second term: to first order, the negative-frequency blocking line has the same departure (but in opposite direction) from the capillary asymptotic limit as the blue horizon (for $T\geq T_c$).

The cusp ($T_c, U_c)=(0.425\mathrm{s},-0.178\mathrm{m.s^{-1}}$) corresponds to the intersection of two saddle-node lines (the white and blue horizons), and is associated to an inflection point of the dispersion relation~\eqref{cubic-dispersion}. In dynamical-systems theory, we anticipate a so-called pitchfork bifurcation~\cite{Poston}: the merging of two saddle-node bifurcations is equivalent to the appearance of a fictive symmetry in the representation space of the cubic-in-$k$ equation~\eqref{cubic-dispersion}. The resulting pitchfork bifurcation is then associated with the breaking of this symmetry, which is absent from the original system. In other words, below $T_c$, the blocking lines or horizons corresponding to both saddle-node bifurcations completely disappear.

Note that the cusp is analogous to a critical point (second-order phase transition) in a thermodynamical phase diagram (Fig.~\ref{Fig:phase}). The saddle-node lines (first-order phase transitions) separate the analogues of thermodynamical phases~\cite{NJP10}. One can distinguish seven regions of interest, marked by Roman numerals, which can be grouped into four phases: $A=I+II+III$, $B=IV$, $C=V+VI$ and $D=VII$.

The vertical line $T=T_c$ roughly separates the capillary ($T<T_c$) and the gravity ($T>T_c$) regimes. The $A$ phase corresponds to a simple root of the cubic dispersion relation, where the incident wave can be of capillary ($I$) or gravity ($II$ or $III$) type. The saddle-node line ending on $U_c$ (critical point) and $U_\gamma$ (tricritical point at infinity) corresponds to the threshold for the simultaneous appearance of blue-shifted waves and capillary waves propagating in the same direction as the incident ones ($B$). The $C$ and $D$ phases are characterized by the presence of negative energy waves. In the $D$ phase, these negative energy waves are of the gravity and capillary type and propagate in the same direction as the counter-flow. The $C$ phase is the forbidden region for gravity waves coming from $A$ across $B$ or $D$. Only gravity waves from the $A$ phase (after mode conversion) or directly capillary waves are allowed to go into the $C$ phase.

\section{Experimental setup}
\begin{figure}[!htbp]
\begin{center}
\includegraphics[width=16cm]{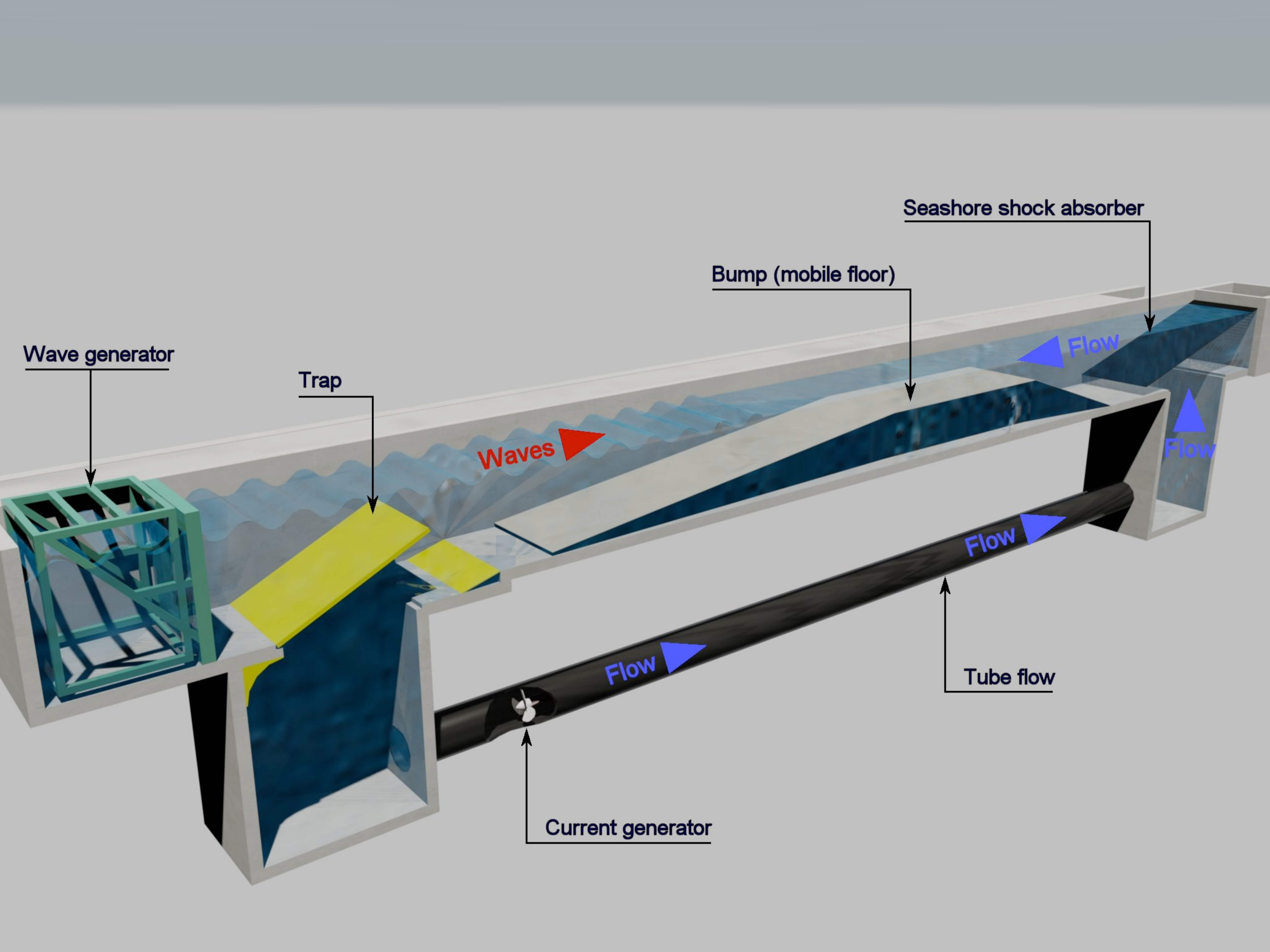}
\caption{Experimental setup. Characteristics of the bump (from left to right): linear slope with angle $\alpha_1=7.5^\circ$ and length $l_1=8m$, flat part $l_2=4.80m$, linear slope $\alpha_3=-18.5^\circ$ and $l_3=3.30m$. Water depth (min--max): 50cm--160cm.}
\label{Fig:ACRI}
\end{center}
\end{figure}

We performed laboratory experiments to corroborate various aspects of the phase diagram in Fig.~\ref{Fig:phase}. These experiments were performed at ACRI, a private research company working on environmental fluid mechanics such as coastal engineering. The experiment features a wave-tank 30m long, 1m80 large and 1m80 deep, see Fig.~\ref{Fig:ACRI}. The piston-type wave-maker can generate waves with periods $T=$0.35-3s and typical wave heights of 0.5-30cm. A current can be created along or opposite to the direction of wave propagation with a maximum flow rate around 1.2m$^3/$s. The waves themselves are recorded using several video cameras and the videos are digitalized and calibrated.

In order to generate a gravity-wave horizon, a bump is immersed into the channel. The bump has a positive and a negative slope separated by a flat section. We send a train of progressive water waves onto the bump, hindered by a reverse fluid flow produced by a pump. 

The background flow velocity depends on the water depth through flow rate conservation. The counter-current accelerates as the water height diminishes, reaches a maximum on the flat part of the bump before slowing down again as the bump height decreases. 

\section{Experimental results}
\label{S:results}
\begin{figure}[!htbp]
\vspace*{-4ex}
\begin{center}
\includegraphics[width=7cm]{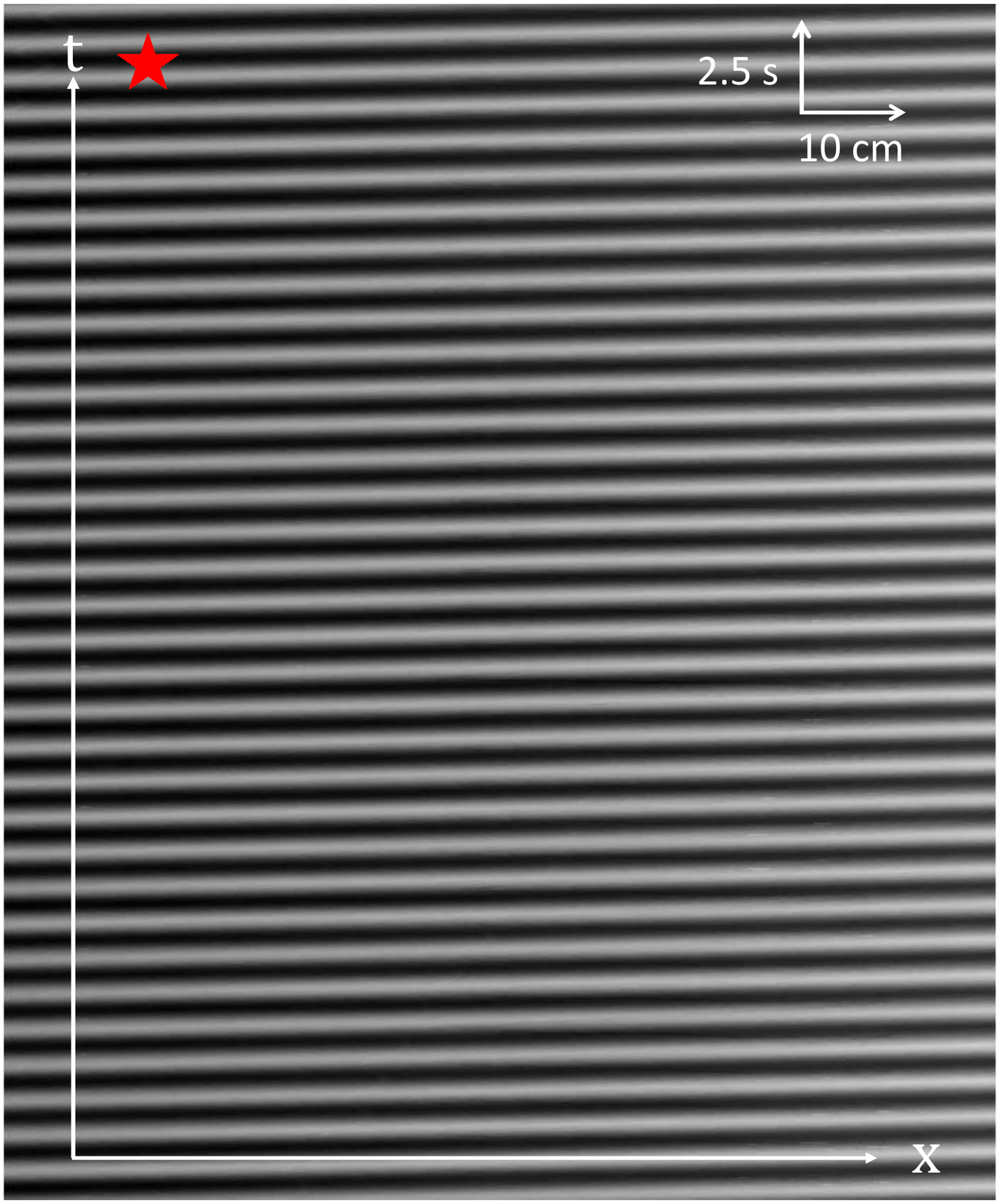}
\includegraphics[width=7cm]{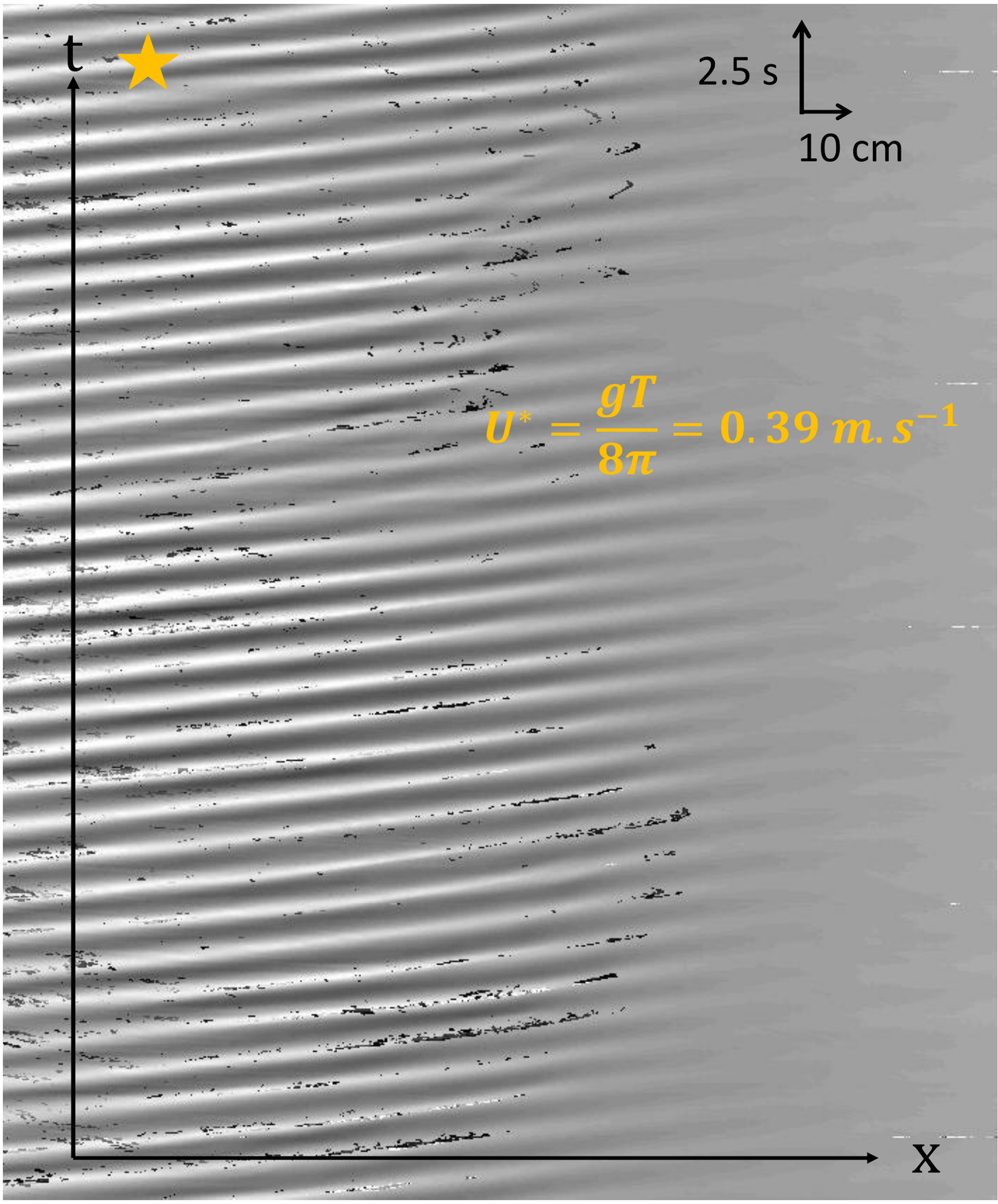}

\vspace*{-4ex}
\includegraphics[width=7cm]{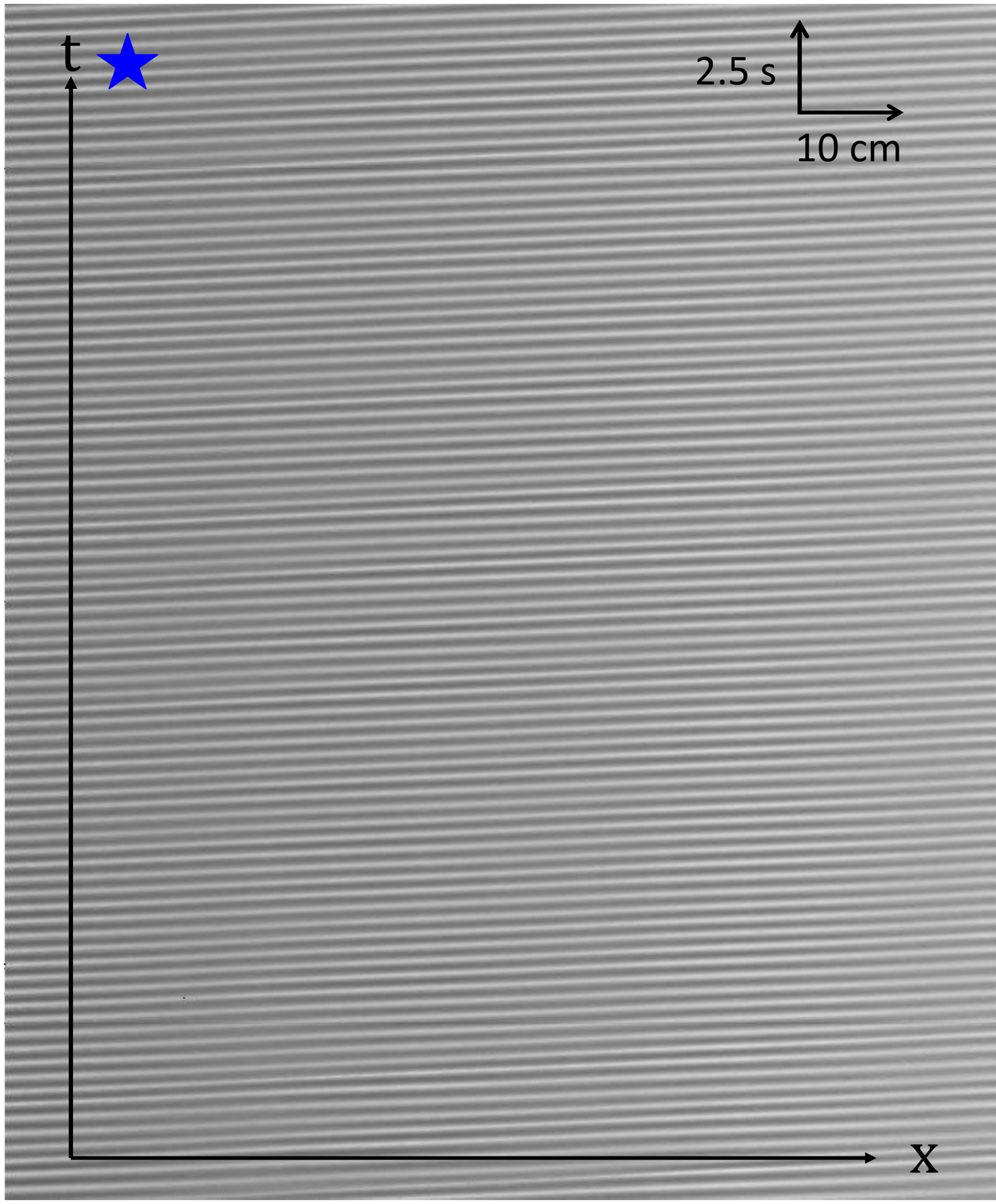}
\includegraphics[width=7cm]{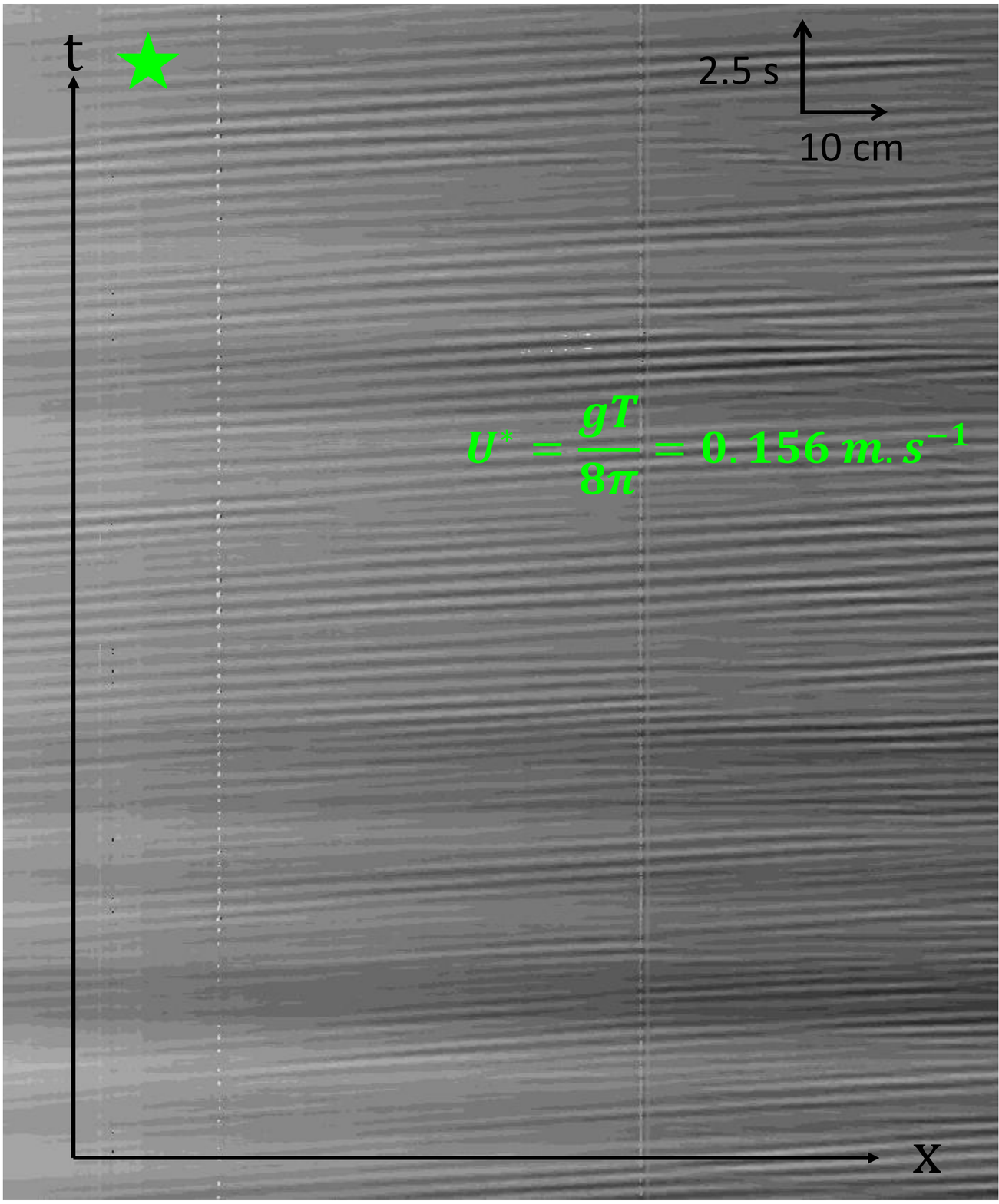}
\vspace*{-2ex}
\caption{Spatio-temporal diagrams for the four values of $(U,T)$ marked with coloured stars in Fig.~\ref{Fig:phase}. The light and dark lines represent the world-lines of crests and troughs, respectively. The diagrams show: the normal propagation of a gravity wave against a moderate counter-current (top left); the blocking of a gravity wave at a blocking line or white-hole horizon due to a strong counter-current (top right); the normal propagation of a capillary-gravity wave in the absence of a counter-current (bottom left); and the propagation of a capillary-gravity wave across a region with a counter-flow velocity well above the gravity-wave blocking value (bottom right). 
The width of the images along the x-coordinate is 78.1cm (top left), 191.7cm (top right), 78.1cm (bottom left) and 113.92cm (bottom right). Other parameters: see main text. 
}
\label{Fig:spatio}
\end{center}
\end{figure}

We have performed detailed measurements at the $(U,T)$ values marked by the four coloured stars in Fig.~\ref{Fig:phase}. Continuous low-amplitude wave-trains were used in order to minimize non-linear effects. The corresponding experimental space-time diagrams are shown in Fig.~\ref{Fig:spatio}.

In the absence of a counter-current, the gravity and capillary terms in the dispersion relation~\eqref{cubic-dispersion} are equal for $\omega= \sqrt{2g/l_c}=86\mathrm{rad.s^{-1}}$ ($f=13.7$Hz), i.e.\ $T=0.073\mathrm{s}$. The range 
$T>0.1s$ corresponds (by convention) to a pure gravity regime, while a pure capillary regime exists for $T<0.04\mathrm{s}$. In the presence of a counter-current, one must look at the phase diagram in Fig.~\ref{Fig:phase} to distinguish the gravity and capillary influence. The upper diagrams in Fig.~\ref{Fig:spatio} correspond to waves with a period
of $T=1\mathrm{s}$. These are therefore pure gravity waves, and we expect them to be blocked near $|U_g|=gT/8\pi=0.39\mathrm{m.s^{-1}}$. The lower diagrams show the propagation of waves with a period $T=0.4\mathrm{s}$, approximately the lowest period allowed by the wave-maker. For weak counter-currents, these behave as pure gravity waves, but they should suffer a strong blueshifting towards the pure capillary regime as the counter-current increases, and penetrate through the gravity-wave white-hole horizon.
The four diagrams in Fig.~\ref{Fig:spatio} thus correspond to the following cases.

(a) First (red star), we recorded the normal propagation of a gravity wave of $T=1\mathrm{s}$ and amplitude $A=3\mathrm{cm}$ against a moderate counter-current ($|U|$ increases from $0.074\mathrm{m.s^{-1}}$ to $0.087\mathrm{m.s^{-1}}$ from left to right, well below $|U_g|=gT/8\pi=0.39\mathrm{m.s^{-1}}$), see Fig.~\ref{Fig:spatio} (top left).

(b) Second (yellow star), still for gravity waves of $T=1\mathrm{s}$ and $A=3\mathrm{cm}$, we depict the range $|U|=0.45$--$0.55\mathrm{m.s^{-1}}$ for which we recover the existence of a white hole marking a forbidden region into which gravity waves cannot enter ($|U_g^\text{exp}|\approx 0.53\mathrm{m.s^{-1}}$), see Fig.~\ref{Fig:spatio} (top right), in agreement with  the value measured in~\cite{Chawla1, Ma2010}. We will come back to the apparent mismatch with the theoretical prediction $|U_g^\text{th}|=0.39\mathrm{m.s^{-1}}$ in the following section.  Note that the white horizon or blocking line is actually blurred into a ``blocking region''. Also notice the clear blue-shifting due to the increasing counter-current: the slope of the world-lines increases from left to right, in full analogy with the behavior of light close to a gravitational fountain. Indeed, the slope is the inverse of the phase velocity and therefore proportional to the wavelength. The slope of incoming rays grows until the rays disappear at the horizon. We point out forcefully that the slope does not increase to infinity. This is due to the dispersive effect close to the horizon, which leads to an Airy regularization mechanism, which we will discuss below. A ``trans-Planckian'' problem is thus avoided for the incident wave, since the blue-shifting does not become infinite, as in the purely relativistic case. However, as we will also discuss below (see  Fig.~\ref{Fig:secondary-transplanckian}), there is still a problem for the mode-converted blue-shifted waves (as well as for the negative waves). In the pure gravity case, these would have $k\to \infty$ as $U\to 0$: if capillarity did not come into play, then the trans-Planckian problem would, in a sense, simply be displaced to flat space-time.

(c) Third (blue star), we show the normal propagation of a capillary-gravity wave ($T=0.4\mathrm{s}$ and $A=1-2\mathrm{cm}$) in the absence of a counter-current ($U=0$), see Fig.~\ref{Fig:spatio} (bottom left). 
Note that there is an excellent agreement between theory and experiment about the wavelengths in this case: $\lambda_\text{exp}=0.25$m versus $\lambda_\text{th}=0.249$m in the whole range $h=0.50\mathrm{m}-1.60\mathrm{m}$.

(d) Finally (green star), we show the propagation of a capillary-gravity wave ($T=0.4\mathrm{s}<T_c$, $A=1$cm) against a counter-current well above the blocking value: $|U|=0.232$--$0.275\mathrm{m.s^{-1}}$ versus $|U_g|=gT/8\pi=0.156\mathrm{m.s^{-1}}$, see Fig.~\ref{Fig:spatio} (bottom right). From the diagram, we conclude that there is a complete penetration across this part of the white-hole region (forbidden for gravity waves), and barely any noticeable blue-shifting, contrarily to the case (b) of the blocked gravity waves. This last diagram therefore shows a double discrepancy with the theoretical expectation. 

\begin{figure}[!htbp]
\begin{center}
\includegraphics[width=10cm]{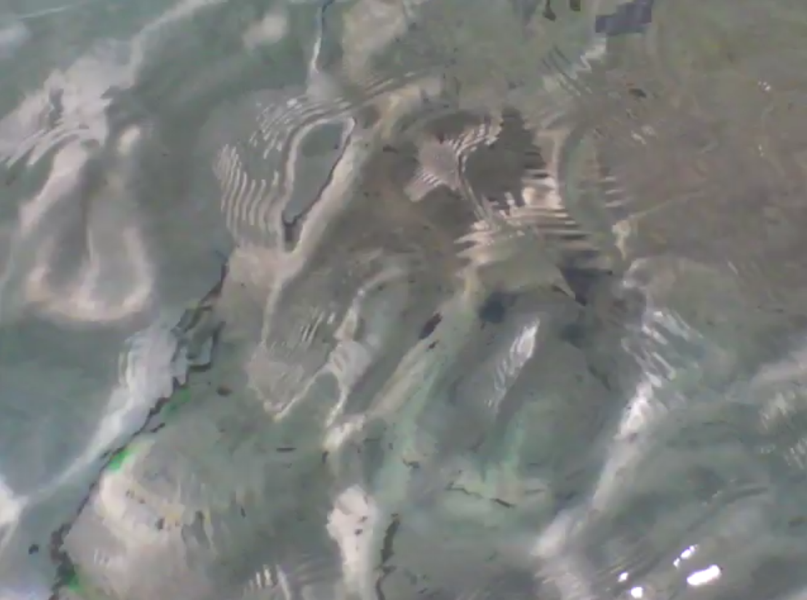}
\caption{Appearance of capillary waves for $T=0.4\mathrm{s}$ in the presence of a strong counter-current $U$.}
\label{Fig:capillary-conversion}
\end{center}
\end{figure}

\begin{figure}[!htbp]
\begin{center}
\includegraphics[width=10cm]{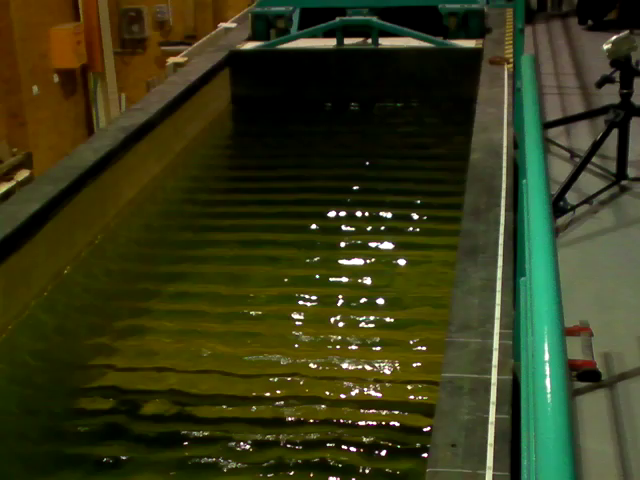}
\caption{Development of transversal instability for $T=0.4\mathrm{s}$: near the wave-maker (at the furthest end of the channel), the wavefronts are nearly perfectly perpendicular to the channel's edges. Towards the bottom of the picture, the wavefronts start to deform under the effect of a transversal instability, and the free water surface acquires a `fish-scale' pattern. The capillary waves in Fig.~\ref{Fig:capillary-conversion} appear on the fronts of these fish scales.}
\label{Fig:transversal-instability}
\end{center}
\end{figure}

First, the lack of any strong blueshifting with respect to the previous case means that the transition to the capillary regime has not been fully completed for these values of $U$. Indeed, for the counter-currents in the range of this camera position ($|U|=0.233-0.275\mathrm{m.s^{-1}}$), $\lambda_\text{th}=8.95-6.13\times 10^{-3}$m whereas the measured $\lambda_\text{exp}$ is of the same order of magnitude as the incident wavelength. The capillary conversion seems to have taken place at a much higher value of the counter-current than expected. Indeed, further upstream (at higher values of the counter-current $|U|$), the appearance of capillary waves can be observed with the naked eye, see Fig.~\ref{Fig:capillary-conversion}. We believe that the mismatch is due to the appearance of a transversal instability, see Fig.~\ref{Fig:transversal-instability}, which blurs the capillary conversion. We plan experiments with a narrower wave channel in the near future in order to reduce this transversal instability and study the conversion to the capillary regime in more detail. 

Second, in the absence of such a full transition to the capillary regime, the waves should have been blocked at (or near) the gravity-wave blocking velocity $|U_g|=gT/8\pi=0.156\mathrm{m.s^{-1}}$. This blocking has not taken place either. This second discrepancy is in the line of the mismatch mentioned above in the pure gravity case, and we will come back to it in the next section. 

\section{Airy interference and gravity-wave blocking}
\begin{figure}[!htbp]
\begin{center}
\includegraphics[width=9cm]{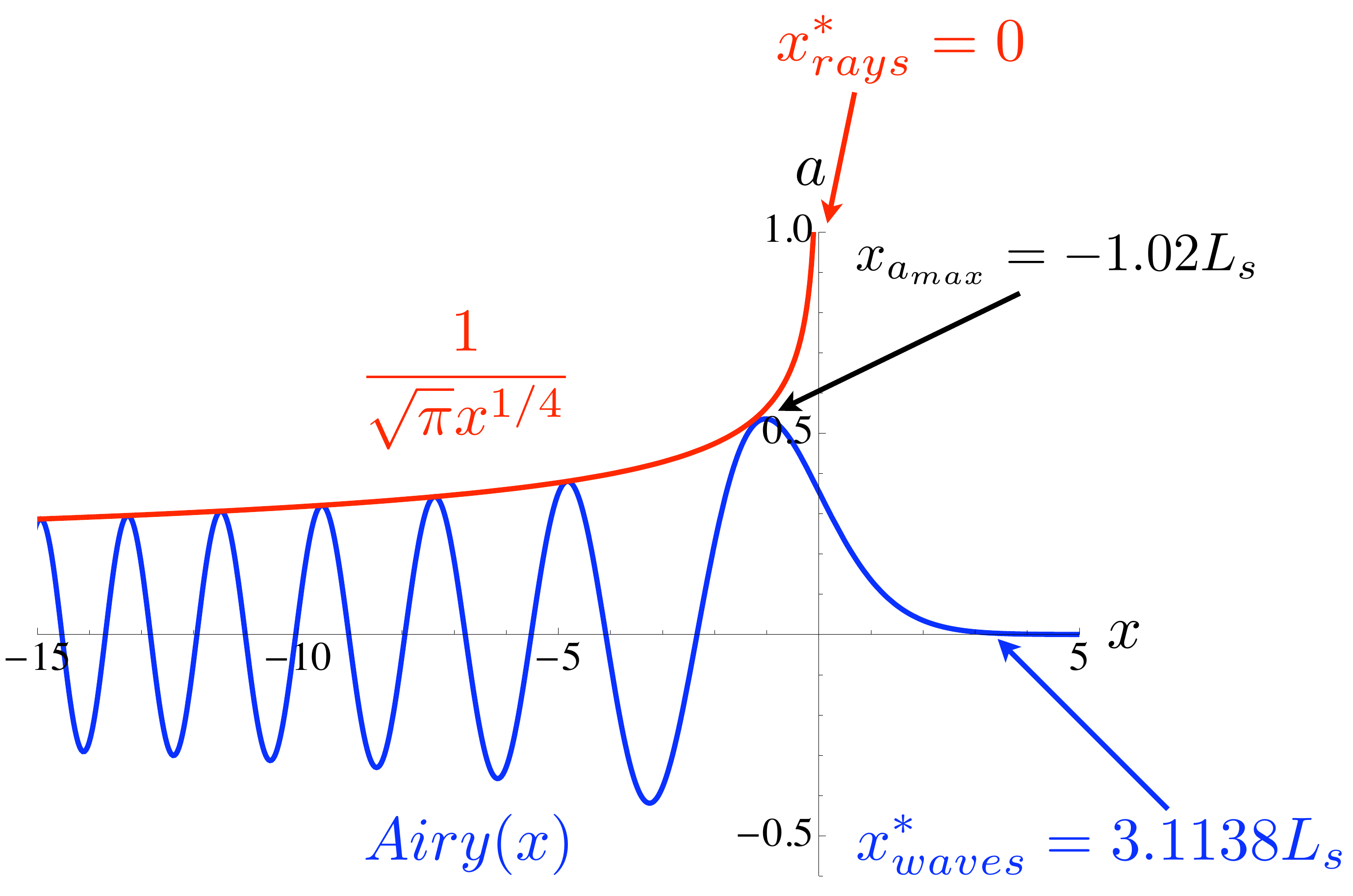}
\caption{Blocking of rays (in red) versus waves (in blue). In the ray approximation, the amplitude theoretically diverges towards the blocking point $x^*_\text{rays}$, beyond which it vanishes discontinuously. This divergence is regularized in the wave picture through Airy interference. The wave blocking point $x^*_\text{waves}$ lies several Airy stopping lengths $L_s$ further than $x^*_\text{rays}$.}
\label{Airy}
\end{center}
\end{figure}

Airy interference provides the crucial mechanism through which a divergence of the amplitude is avoided~\cite{PRL09} in the regime $kh\gg 1$, see Fig.~\ref{Airy}. An explicit expression for the stopping length $L_s$ associated to the Airy interference in the case of pure gravity waves is 
\begin{equation}\label{stopping-length}
L_s = \frac{1}{16(2\pi^5)^{1/3}} gT^{5/3}\left(\frac{dU}{dx}\right)_{x=x_*}^{-1/3},
\end{equation}
with $x_*$ the horizontal blocking position. A simple derivation of this expression can be found in the Appendix. Because of the geometry of our experiment, the background surface velocity evolves linearly: $|U(x)|=0.51-0.05x$, where $x=0$ corresponds to the kink in the mobile floor (note that the $x$-axis is oriented along the background flow, i.e.\ from right to left in Fig.~\ref{Fig:ACRI}). From Fig.~\ref{Airy}, the apparent mismatch between the well-known theoretical prediction $U_g^\text{th}=-\frac{gT}{8\pi}$ for the counter-flow velocity at blocking ($|U_g^\text{th}|=0.39\mathrm{m.s^{-1}}$ for $T=1\mathrm{s}$), and the measured value ($|U_g^\text{exp}|=0.53\mathrm{m.s^{-1}}$) can also partially be understood. $U_g^\text{th}$ is obtained in the ray approximation. The waves will actually be blocked a certain distance $\Delta x^*_\text{waves}$ further due to the
Airy interference. The experimental blocking position $x^\text{exp}_*=-0.36$m corresponds to the wave-blocking position. 
Taking the conservative assumption that a wave is considered ``blocked'' (i.e., it is no longer detected on camera) when its amplitude has decreased below 1\% of its maximum, one obtains $\Delta x^*_\text{waves}=3.11L_s$, i.e.\ (for $T=1\mathrm{s}$) $\Delta x^*_\text{waves}=0.61$m. The ray-divergence position would then be at $x^\text{exp}_*+\Delta x^*_\text{waves}=0.25$m, corresponding to a velocity $U_g^\text{exp-ray}=0.50\mathrm{m.s^{-1}}$. The theoretical blocking position corresponding to $|U_g^\text{th}|=0.39\mathrm{m.s^{-1}}$ is $x_*^\text{th}=+2.42m$. In other words, Airy interference explains roughly 20--25\% of the difference between the theoretical prediction and the experimental measure.

\begin{figure}[!htbp]
\begin{center}
\includegraphics[width=6cm]{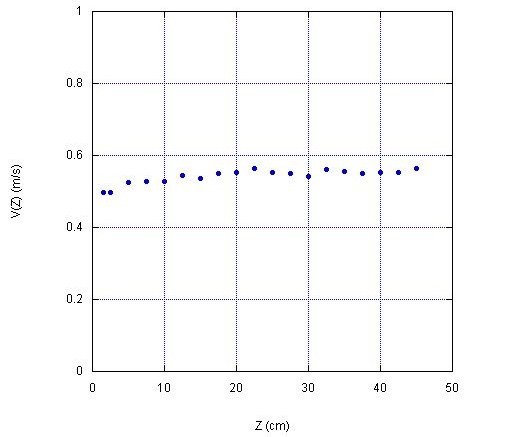}
\includegraphics[width=6cm]{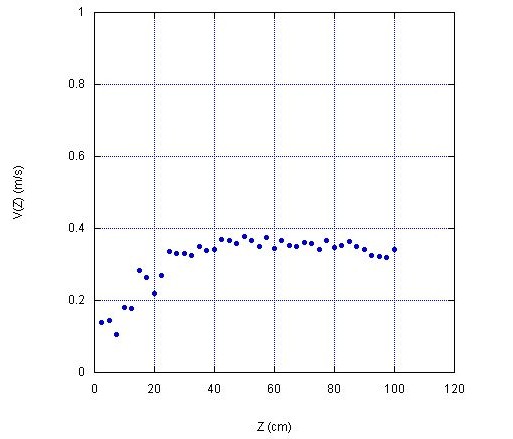}
\caption{Vertical background-flow velocity profiles: nearly plug-type profile on the bump at \mbox{$x$=-2.43m} (left) and parabolic profile towards the wave-maker at $x$=4.20m (right), where $x$=0 corresponds to the kink in the bump, see Fig.~\ref{Fig:ACRI}.}
\label{Fig:vertical-velocity}
\end{center}
\end{figure}

A second element which contributes to the difference between $U_g^\text{th}$ and $U_g^\text{exp}$ lies in the decrease with depth of the real velocity profile. The theoretical prediction for the value at the surface should, in a real experiment, be considered as an averaged (integrated) value over some depth, necessarily leading to a slightly higher value at the surface. The vertical velocity profile is approximately of the so-called plug type on the flat part of the bump and acquires a parabolic form after the flow has decelerated on the descending slope of the bump, see Fig.~\ref{Fig:vertical-velocity}. Comparison of the surface values with the average, vertically integrated values gives differences of 5--20\%. 
Interpolating between the cases represented in Fig.~\ref{Fig:vertical-velocity}, we obtain an estimated difference of $\sim 10\%$ between the theoretically predicted value and the value measured experimentally at the surface near the blocking point ($x_*^\text{exp}=-0.36$m for $T=1\mathrm{s}$). 

A third element which might be thought to be important is that the blocking velocity $U_g^\text{th}=-\frac{gT}{8\pi}$ is obtained in the pure gravity-wave limit, whereas even for $T=1$s, a small capillary influence persists and slightly increases the real blocking velocity. However, this difference is negligible, as can be seen from Fig.~\ref{Fig:spatio}, where it corresponds to the departure between the red line $U_g$ and the corresponding black line $U^*$ obtained numerically from the full Eq.~\eqref{quintic}. It can be estimated quantitatively as follows. In the regime $kl_c \ll1$, the dispersion relation~\eqref{cubic-dispersion} can be approximated by
\begin{equation}
\omega \simeq Uk+\sqrt{gk}\left(1+\frac{(kl_c)^2}{2}\right)
\end{equation}
The condition $\frac{d\omega}{dk}=0$ for wave blocking becomes
\begin{equation}
U^*=-\sqrt{\frac{g}{4k}}(1+\frac{5}{2}l_c ^2k^{2})
\end{equation}
In the limit $l_c=0$, the blocking wavenumber $k_g$ (corresponding to $U_g=-g/4\omega$) gives $k_g=\frac{4\omega^2}{g}$. Inserting this value in the previous expression leads to
\begin{equation}
U^*\simeq U_g+\Delta U^*=-\frac{gT}{8\pi}-80\pi^3 \frac{l_c ^2}{gT^3}
\end{equation}
For $T=1s$, $\Delta U^* \simeq -0.002{\mathrm m.s^{-1}} \ll U_g$. 

Non-linear effects could also play an important role, in spite of our attempts to limit them by working with small amplitudes. We limit ourselves to two comments. First, the finite wave amplitude $A$ increases the blocking velocity. This well-known (but poorly understood) phenomenon~\cite{Chawla1,Suastika1} can to a first approximation be modelled as an effective surface tension (see e.g.~\cite{lamb}):
\begin{equation}
 \omega^2=gk(1+A^2k^2)
\end{equation}
(for pure gravity waves in the absence of a counter-current). Since $A/l_c\sim 10$ in our experiments, it is clear that the finite amplitude has a much stronger influence on the blocking velocity than the intrinsic surface tension, and is perhaps the main contributor to the difference between $U_g^\text{th}$ and $U_g^\text{exp}$. This illustrates the importance of generating waves with a low factor $Ak$ ($Ak\sim 0.1$ at $T=1$s in our current setup). A second non-linear effect which plays an important role in certain other wave-blocking experiments~\cite{Chawla2,Ma2010} is the Benjamin-Feir instability, which leads to the appearance of so-called side-bands (excitations at frequencies slightly different from the fundamental one). However, we have verified conservation of period in our experiments, thereby nearly excluding this possibility.

We believe the above elements to provide a reasonable explanation for the apparent mismatch between $|U_g^\text{exp}|\approx 0.53\mathrm{m.s^{-1}}$ and $|U_g^\text{th}|=0.39\mathrm{m.s^{-1}}$. It should be noted that this mismatch is well known in the fluid-mechanics community, but not well understood. For example, \cite{Chawla1} and \cite{Ma2010} also observe the blocking of $T=1$s waves at a countercurrent velocity $|U_g^\text{exp}|\approx 0.53\mathrm{m.s^{-1}}$, but do not attempt to interpret this discrepancy with the theoretical prediction.

Also note that similar arguments would apply to the case of the gravity-capillary waves at $T=0.4\mathrm{s}$. There, however, the additional appearance of a transversal instability mentioned above further complicates matters and further experiments are required to clarify the situation. 

It is remarkable that the Airy interference hybridizes the character of the incoming wave. A hybrid is created between the original wave (through the period $T$, in the expression~\eqref{stopping-length} of the stopping length $L_s$) and the background flow (through $\frac{dU}{dx}$). In our experiment, we are sending continuous wave-trains. If one were to send wave-packets (``particles'', i.e.: superpositions of waves), then these would be deformed into superpositions of wave-flow hybrids, or ``hybridons''.
As a matter of fact, we can take this observation further. The Airy stopping length obeys $L_s \propto \lambda^*\mathscr{U}_n^{1/3}$, where $\lambda^*$ is the wavelength at blocking, and $\mathscr{U}_n$ the dimensionless ``Unruh'' number $\mathscr{U}_n=\omega\left(\frac{dU}{dx}\right)^{-1}_{x_*}$ obtained from the two characteristic ``frequencies'' at blocking: the wave frequency $\omega$ and the flow gradient $\frac{dU}{dx}$. This leads to the following interpretation. Dispersion has a double role in the near-horizon physics. It keeps the wavenumber finite (i.e., it solves the trans-Planckian problem---see next section), thereby avoiding the  first relativistic ray-theory pathology \mbox{$\lambda^*\to 0$}. Second, it creates an interference mechanism which hybridizes this wavelength with the background flow by modulating it through $\mathscr{U}_n$. Dispersion thus replaces the wavelength by a characteristic interference length $L_s$. This mechanism of interferences allows to solve the second pathology associated to the ray theory: the infinite amplitude at the blocking point. Indeed, when $\mathscr{U}_n\gg 1$, i.e.\ when the frequency of the wave is large compared to the spatial variation of the background flow velocity, then the WKB-approximation is valid. Note that $\mathscr{U}_n\gg 1$ also leads to $\exp\left(\omega/\left(\frac{dU}{dx}\right)_{x_*}\right)-1\gg 1$ and therefore to negligible Hawking radiation. Near the blocking point, though, one always has $\mathscr{U}_n\sim 1$. The WKB approximation then breaks down, and two resonance mechanisms come into play. The first one (Airy interference) is an adiabatic process and the second one (Hawking radiation) is a non-adiabatic process, see the Chapter on ``The Basics of Water Waves Theory for Analogue Gravity'' elsewhere in this Volume.

As a final note, we should point out that we have neglected the presence of a zero mode in our considerations on the Airy mechanism. Such a zero mode (an $\omega=0$ solution to the dispersion relation, or superposition of various such solutions) would deform the free surface and complicate the interference pattern, see~\cite{Silke,807309} and the discussion in the Chapter ``The Cerenkov effect revisited: from swimming ducks to zero modes in gravitational analogs'' elsewhere in this Volume. This omission is justified since in the regime $kh\gg 1$ one can minimize the amplitude of the zero mode by working at low velocities and limiting the slope of the bump. Note that $\gamma\neq 0$ implies the existence of a threshold $|U|\geq |U_\gamma|$ for the appearance of a zero mode, contrarily to the pure gravity case.

\section{The trans-Planckian problem}
\label{S:transplanckian}
Our experimental results to corroborate the theory developed in~\cite{NJP10} have been slightly marred by the appearance of a transversal instability, which we hope to remedy using a narrower wave-channel. Nevertheless, there is little doubt that capillary waves can penetrate through a gravity-wave blocking line. The full strength of this statement becomes clear in the context of the gravitational analogy. The surface tension constitutes a high-$k$ dispersive correction to the low-$k$ gravity-wave theory. Such dispersive corrections can therefore completely alter the properties of a horizon: dispersive horizons are no longer one-way membranes, and the infinite blueshifting associated with strictly relativistic horizons disappears. The example of surface waves shows that this statement can be true even if the dispersion is (initially) subluminal.

\begin{figure}[!htbp]
\begin{center}
\includegraphics[width=10cm]{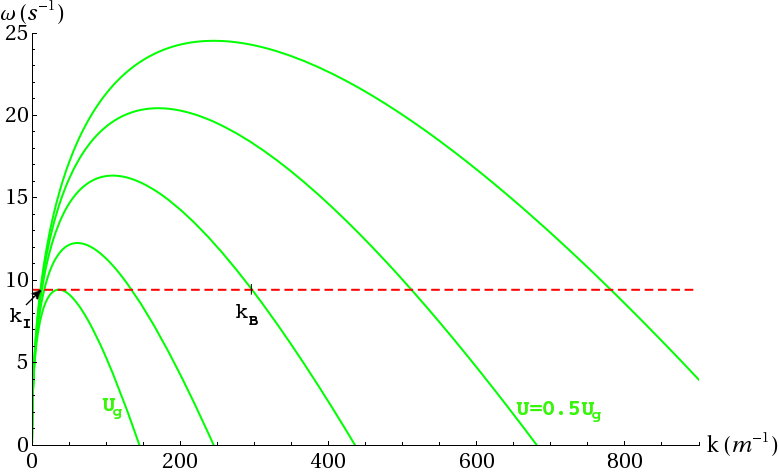}
\caption{Gravity waves and the trans-Planckian problem. For a given frequency $\omega$ (dashed red line), a blue-shifted wave $k_B$ is created from the incident wave $k_I$ through mode conversion at the blocking line $U=U_g$. Since $k_B$ has $c_g<0$, it moves away from the horizon towards lower $|U|$. As $|U|\to 0$, $k_B\to \infty$, leading to a secondary trans-Planckian problem in flat space-time.}
\label{Fig:secondary-transplanckian}
\end{center}
\end{figure}

The idea that dispersive corrections could solve the trans-Planckian problem of gravity has from the start been one of the cornerstones of the  analogue gravity programme~\cite{Unruh}. Most work has historically focused on the study of subluminal (``normal'') dispersion. This seems curious in the light of the following observations. In the pure gravity-wave (subluminal) regime, the trans-Planckian problem is indeed avoided for the incident wave: it is mode-converted into a blue-shifted wave and the wave bounces away from the horizon (in terms of the group velocity; the phase velocity is still directed towards the white hole). However, as this blue-shifted wave approaches ``flat space-time'' (i.e., as the counter-current velocity $|U|\to 0$), a new, secondary trans-Planckian problem arises: the wavenumber of the blue-shifted wave $k_B\to \infty$, see Fig.~\ref{Fig:secondary-transplanckian}. This was observed earlier~\cite{Jacobson:1996zs} with respect to Unruh's original subluminal model~\cite{Unruh:1994je}, and related problems with other subluminal models were also discussed in~\cite{Corley:1996ar} and~\cite{Jacobson:1999zk}. The same secondary trans-Planckian problem occurs for the negative-frequency waves associated to a Hawking-like process. Actually, even without invoking any horizon effects, a similar problem arises: Any counter-current, no matter how small, would allow for the existence of both blue-shifted and negative-energy waves with infinitely small wavelengths. Purely subluminal dispersion would then solve the primary trans-Planckian problem associated to the horizon, at the cost of creating a new one in flat space-time.

More complicated dispersion relations, e.g.\ as in Helium-II (superfluid $^4$He), which has a ``roton'' minimum in the $\omega'-$vs$-k$ diagram, could overcome this problem by letting the outgoing blue-shifted wave decay at the end of the quasiparticle spectrum into two rotons, with the same total energy and momentum (i.e., $\omega\rightarrow \omega/2 + \omega/2;~ k\rightarrow k/2 + k/2$)~\cite{Jacobson:1996zs}. The secondary trans-Planckian problem is then avoided because the dispersion curve ends in such a two-roton decay channel.\footnote{Actually, this is not the end of the story: these rotons are still subject to the background flow, and will therefore also start blueshifting, just like the original mode, and again split into two rotons each at the end of the quasiparticle dispersion curve, etcetera, leading to an apparently endless creation of rotons. This process is limited because the roton creation will deplete the superfluid component, and ultimately destabilizes the white-hole configuration. Also note that the rotons will relax after some time due to interaction with the environment and condense into a roton BEC~\cite{iordanskii}. If one takes such a $^4$He-like dispersion model seriously for true gravity, this could lead to the creation of a photon condensate near a white hole, or vice versa: outgoing particles from a black hole might originate from a condensate in curved spacetime. Although such a scenario might not be as crazy as it sounds~\cite{blhu}, we will not pursue this exotic line of thought further here, and stick to ``simpler'' solutions of the trans-Planckian problem.}

Another obvious way of resolving the secondary trans-Planckian problem with subluminal dispersion is through dissipation, for example due to viscosity. Note that viscosity, apart from leading to dissipation, necessarily also introduces dispersion~\cite{Visser:1997ux}. Dissipation in media is quite generic, often unavoidable (as in optical media, where dissipation and dispersion are coupled through the Kramers-Kronig relations), and might be relevant for our ``fundamental'' space-time as well. Actually, from a theoretical QFT point of view, Lorentz symmetry violation is automatically accompanied by dissipation under quite general assumptions, and dissipative effects should therefore in principle be treated together with dispersive ones~\cite{Parentani:2007uq}. However, dissipation is formally much harder to treat than dispersion, and it has received little attention in the context of (analogue) gravity. Also, the relevance of dissipation obviously depends on its characteristic scale. Here, we mainly wish to stress that the secondary trans-Planckian problem in surface waves is solved through dispersion before dissipation becomes relevant.

These observations suggest a more general interpretation for our results with water waves. The mesoscopic scale of surface tension ``saves'' the fluid continuum approximation from breaking down in the presence of a counter-current: The capillary behaviour at high $k$ is essential in order to avoid a trans-Planckian pathology. We therefore expect that any system displaying analogue gravity behaviour at low $k$ through propagation on a moving background medium will necessarily have superluminal dispersion at sufficiently high $k$'s, unless dissipation kills the whole phenomenon before such scales are actually reached. The exotic case of Helium-II mentioned above might be the exception that confirms the rule, since even then, the dispersion relation is superluminal for a certain range of high wavenumbers beyond the roton minimum.

Moreover, we can establish the following general rules. If the first corrections at intermediate $k$ are subluminal, followed by superluminal corrections at high $k$, as in the case of deep-water waves, then there will always be a white horizon and a blue horizon, leading to the possibility of a Badulin-type double-bouncing scenario. There will then also always exist some counter-current velocity $U_c$ for which these white and blue saddle-node points merge into a pitchfork bifurcation and both horizons disappear, allowing for direct dispersive penetration. Such direct dispersive penetration is actually even more universal: it suffices to have a superluminal correction at high $k$ to a relativistic low-$k$ behaviour, irrespective of the intermediate regime. This is true, e.g., for phonons in BECs~\cite{Pitaevskii}, just like for capillary surface waves: sufficiently blueshifted   (``superblueshifted'') modes will always be able to penetrate through any counter-flow barrier, unless dissipation prevents such superblueshifting. The case of surface waves is in a sense richer than that of BEC-phonons, in that there is a true blocking line for low-frequency gravity waves, which cannot directly penetrate the horizon. For BECs, the absence of an intermediate subluminal correction implies that even low-frequency phonons can in principle superblueshift and cross the horizon directly. Finally, in spite of our several comments regarding dissipation, it seems that both horizon-crossing scenarios can indeed be fully realized for surface waves before being dissipated.

To sum up, to enter a white hole---or, by time-inversion: to escape a black hole---one has to either tune the period to be subcritical, or bounce on two horizons.

The bottom line is of course what this teaches us for real gravity. Here the issue is more complicated, because extrapolation from the current state of observations has so far not given any evidence for dispersion (or dissipation) even at the Planck scale. Other complications might also arise which are peculiar to real gravity. For example,  in~\cite{Barbado:2011ai} it was shown using a simple toy model that superluminal dispersion would render gravitational black holes strongly unstable, due to the leaking of resonant modes. In any case, we may conclude that, if dispersion is indeed relevant in gravity (possibly even far beyond the Planck scale), then subluminal dispersion alone would most certainly not be sufficient to solve the trans-Planckian problem.

\section*{Appendix: Airy stopping length}
Smith was the first to derive the Airy equation in 1975 by performing an asymptotic expansion of the water-wave equations (Euler equations + continuity equation + boundary conditions) close to the caustic~\cite{Smith}. He inferred the so-called amplitude equation which is a nonlinear Schr\"odinger equation with a term proportional to the distance to the caustic. When the cubic term is negligible, the amplitude equation reduces to the Airy equation. 

Following Smith and after tedious algebra (matched asymptotics and WKB solutions), Trulsen and Mei computed the following stopping length (for $\gamma=0$) in 1993~\cite{TM}:
\begin{equation}
L_s=\left(\frac{U_*^2}{2 k_* \omega \left(\frac{dU}{dx} \right)_{x_*}}\right)^{1/3}
\end{equation}

In 1977, Basovich \& Talanov~\cite{Basovich} provided another derivation of the Airy equation by noticing that $\frac{dU}{dk}=0$ at the blocking point. Taylor-expanding the function $U(k)$ close to its parabolic minimum and the function $U(x)$ close to the stopping point $x_*$ and taking the inverse Fourier transformation, they deduced the Airy function and the associated stopping length:
\begin{equation}
L_s=\left(\frac{\omega}{4 k_*^3 \left(\frac{dU}{dx} \right)_{x_*}}\right)^{1/3}
\end{equation}

In 1979, Peregrine \& Smith~\cite{PS} used an operator expansion method: the idea is to inverse Fourier-transform a truncated series expansion of the dispersion relation written in the form $G(\omega , k, x)=0$. Again, the cubic Schr\"odinger equation with a spatial term was derived with another expression for the stopping length:
\begin{equation}
L_s=\left(\frac{G_{kk}}{2G_x}\right)^{1/3}
\end{equation}
where the subscripts mean partial derivative and the derivatives are taken at the blocking line. The method was generalized in 2004 by Suastika~\cite{Suastika1, Suastika2} to include viscous dissipation and wave breaking. In 2003, Lavrenov~\cite{Igor} applied a saddle-point method to the Maslov integral representation of the uniform wave field asymptotics in the vicinity of the blocking line. He found:
\begin{equation}
L_s=\left(\frac{\Omega _{kk}}{2\Omega _x}\right)^{1/3}
\end{equation}
where $\omega = \Omega (k, x)$ is the dispersion relation function and the derivatives are taken at the caustic.

We will show that it is possible to derive the stopping length in a very simple fashion, inspired by the method of Basovich \& Talanov, and derive a previously unnoticed scaling law for $L_s$.

We write the background flow velocity near the critical value $U_*=-\frac{gT}{8\pi}$ as a function of $x$ and $k$, and develop both to lowest non-zero order around the stopping length:
\begin{align}
U(k)&\simeq U(k_*)+\frac{U''(k_*)}{2}(k-k_*)^2
= U_*-\frac{U_*^3}{4\omega ^2}(k-k_*)^2\\
U(x)&\simeq U_*+\left( \frac{dU}{dx} \right)_{x_*}(x-x_*)\label{U-of-x}
\end{align}
Equating both into 
\begin{equation}
\frac{U_*^3}{4\omega ^2}(k-k_*)^2+\left( \frac{dU}{dx} \right)_{x_*}(x-x_*)\simeq 0
\end{equation}
and making the substitution
$H(x) \simeq e^{i(k-k_*)x}$
immediately leads to the Airy differential equation
\begin{equation}
\frac{d^2 H}{dX ^2}-XH =0
\end{equation}
where $X=\frac{x-x_*}{L_s}$, with $L_s=\frac{|U_*|}{\left(4\omega ^2 \left( \frac{dU}{dx} \right)_{x_*}\right)^{1/3}}$. 
Thus, $H(x)$ is an Airy function
\begin{equation}
H(x) \simeq Ai \left( \frac{x-x_*}{L_s} \right)=  \frac{1}{\pi}\int _0 ^\infty cos \left ( \frac{1}{3}t^3+\frac{x-x_*}{L_s} t \right)dt
\end{equation}
and 
\begin{equation}
L_s = \frac{1}{16(2\pi^5)^{1/3}} gT^{5/3}\left(\frac{dU}{dx}\right)_{x=x_*}^{-1/3} 
\end{equation}
is the Airy stopping length, which depends both on the incident wave and the background flow: it scales with the period $T$ of the incident wave as $L_s \propto T^{5/3}$ and with the background flow acceleration as $L_s \propto \left(\frac{dU}{dx}\right)_{x=x_*}^{-1/3}$. A straightforward dimensional analysis ($L_s\simeq g^\alpha T^\beta \left(\frac{dU}{dx}\right)_{x=x_*}^{\gamma}$) would only lead to 
$\alpha =1$ and $\beta -\gamma =2$. 

Note that Airy interference requires the flow gradient to remain approximately constant over the characteristic length of the interference process. This can easily be seen in our derivation: Eq.~\eqref{U-of-x} is only a good approximation if $\frac{d^2U}{dx^2}\approx 0$ for $x-x_*=\mathcal{O}(L_s)$. In our experiments, $\frac{dU}{dx}$ is a constant on the whole linear slope of the bump where the horizon $x_*$ is located, so we do not need to worry about this issue.

\subsection*{Acknowledgements}
The authors thank C.~Barcel\'o, D.~Faccio, L.~J.~Garay, Th.~G.~Philbin and G.~E.~Volovik for useful discussions and comments. GR is grateful to Conseil G\'en\'eral 06 and r\'egion PACA (HYDRO Project) for financial support.

\end{document}